\documentclass{aa}  
\usepackage[dvips]{graphicx}  
\usepackage{txfonts}
\usepackage{lscape}
\newcommand{\nc}{\newcommand}
\nc{\Porb}{$P_{\rm orb}$\,}
\nc{\Teff}{T$_{\rm eff}$\,}
\nc{\logg}{log\,$g$\,}
\nc{\kms}{\,${\rm km.s}^{-1}$\,}
\nc{\Msun}{M$_{\odot}\ $}
\nc{\Mcz}{M$_{CZ}\ $}
\nc{\vsini}{$\upsilon \sin i$}
\nc{\vmic}{$\upsilon_{\rm mic}$}
\nc{\vrad}{$v_{\rm rad}$}
\nc{\ALi}{A$_{\rm Li}$\,}
\nc{\Ali}{A$_{\rm Li}$\,}

\begin{document}

\title{Lithium abundances and  extra mixing processes in evolved stars  of M67
\thanks{Based on observations
collected at ESO, Paranal, Chile (VLT/FLAMES program ID 072.D-0309 and
074.D-0179).}$^,$\thanks{Table 6 is only available in electronic form at {\tt http://www.edpsciences.org}.}}

\author{B. L. Canto Martins \inst{1,2},
A. L\`ebre \inst{2},
A. Palacios \inst{2},
P. de Laverny \inst{3},
O. Richard \inst{2}
C. H. F. Melo \inst{4},\\
J. D. do Nascimento Jr \inst{1},
\and J. R. De Medeiros \inst{1}
}
\offprints{B. L. Canto Martins}

\institute{Universidade Federal do Rio Grande do Norte, Departamento de F\'{\i}sica, 59072-970
Natal, RN, Brazil \\ \email{brunocanto@dfte.ufrn.br}
\and
 Universit\'e Montpellier II, CNRS, UMR 5024, Groupe de Recherche en Astronomie et Astrophysique du
Languedoc, Place E. Bataillon, 34095 Montpellier, France
\and
Universit\'e Nice Sophia Antipolis, CNRS, UMR 6202, Observatoire de la C\^ote d'Azur, Laboratoire Cassiop\'ee, 
BP 4229, 06304 Nice, France
\and
European Southern Observatory,
Garching, Germany
}

\date{Received May 2010, Accepted December 2010}

\authorrunning{Canto Martins et al.}
\titlerunning{Li abundance in M67}

 
 \abstract
   {}
   {We present a spectroscopic analysis of a sample of evolved stars  in
  M67 (turn-off, subgiant and giant stars) in order to bring observational
  constraints to evolutionary models taking into account non-standard 
  transport processes.}
   {We determined the stellar parameters (\Teff, \logg, $ [Fe/H]$),  microturbulent and rotational velocities 
and, Lithium abundances (\Ali) for 27 evolved
  stars of M67 with the spectral synthesis method based on  MARCS model
  atmospheres.  We also computed non-standard stellar evolution
  models, taking into account atomic diffusion and rotation-induced
  transport of angular momentum and chemicals that were compared with this set of homogeneous data.}
   { The lithium abundances that we derive for the 27  stars in our
  sample follow a clear evolutionary pattern ranging from the turn-off to
  the Red Giant Branch. Our abundance determination confirms the well known
  decrease of lithium content for evolved stars. For the first time, we
  provide a consistent interpretation of both the surface
  rotation velocity and of the lithium abundance patterns observed in an
  homogeneous sample of TO
  and evolved stars of M67. We show that the lithium evolution is determined
  by the evolution of the angular momentum through
  rotation-induced mixing in low-mass stars, in particular for those with initial masses larger than 1.30
  M$_\odot$ when at solar metallicity.}
   {}



\maketitle


\section{Introduction}

The investigation of the lithium (Li) abundance in different stellar sites
- such as open clusters - is a key element in the study of the chemical
evolution of the Galaxy (D'Antona \& Matteucci 1991). Regarding the formation of 
light elements, unlike beryllium (Be) and boron (B), which are formed only via spallation reactions involving 
protons or alpha particles and atoms of carbon, nitrogen, and oxygen (Reeves, Fowler \& Hoyle 1970), 
lithium is mostly produced during Big Bang nucleosynthesis.
Because they are easily destroyed at low temperature in deep stellar layers, these light elements and especially Li
provide strong constraints to test transport mechanisms in stellar interiors.\\

According to standard stellar evolution, where no mechanisms for the
transport of matter is included in the radiative regions, the only episode
modifying the surface abundance pattern of low-mass stars during their ascent of the Red Giant Branch (RGB) 
 is the first dredge-up (1$^{st}$ DUP) occuring at the base of the RGB. At this stage the convective
envelope deepens and the ashes of hydrogen burning, mainly CN-cycle products, are
dredged-up to the stellar surface, causing the decrease of the
$^{12}$C/$^{13}$C and $^{12}$C/$^{14}$N surface ratios. The regions reached
by the convective envelope are also lithium free and the surface lithium abundance therefore drops
significantly. The importance of these variations depends on metallicity and
initial mass.\\

However, the large number of abundance anomalies detected in main--sequence
and giant stars are evidence pointing to one or more
processes of extra-mixing occurring in the stellar interiors (Pinsonneault, 1997). Several extra-mixing processes were suggested with the
intention of explaining the Li depletion in F and G type stars: microscopic diffusion (Michaud 1986; Chaboyer et al. 1995),
meridional circulation and hydrodynamical turbulent instabilities (Schatzman \& Baglin 1991;
Pinsonneault et al. 1991; Deliyannis \& Pinsonneault 1997 ; Talon \&
Charbonnel 1998; Palacios et al. 2003), internal gravity waves
(Garc\'ia L\'opez \& Spruit 1991; Montalban \& Schatzmann 2000; Talon \&
Charbonnel 2003). However, the physical parameters that control this extra-mixing
are subjects of debate (Charbonnel \& Balachandran 2000). 
Among the candidates, rotation-induced mixing alone (Palacios et al. 2003)
or associated to internal gravity waves (Talon \&
Charbonnel 2003) appears to explain the hot and the cold side of the Li-dip observed for 
main-sequence stars in open clusters. Going beyond the turn-off, Palacios et al.
(\cite{Palacios03}) showed that the same models that reproduce the hot side of
the Li-dip lead to the observed dispersion of Li abundances in subgiants of
open clusters and of the galactic field. Concerning the RGB, abundance variations observed at the surface of stars beyond the 1$^{st}$ DUP indicate the existence of extra mixing processes at this stage (Recio-Blanco \& de Laverny, 2007). However, rotation-induced
mixing does not seem to play a major role in more advanced evolutionary stages, when the
rotation decreases because of the increase of the stellar radius, as shown in
Palacios et al. (\cite{PCTS06}) for globular cluster RGB
stars. Finally let us mention that the second and third dredge-up episodes
that occur during the Asymptotic Giant Branch phase once again modify the
surface abundance patterns, and that in these phases abundance
anomalies also indicate the action of an extra-mixing process that is not yet clearly identified. Li depletion also seems to be 
related to the stellar age (Herbig 1965; Duncan 1981; Soderblom 1983; Fekel
and Balachandran 1993), and it is also sometimes considered to be
connected with other parameters such as metallicity, activity, and mass
(Pallavicini et al. 1987; Spite \& Spite 1982; Randich et al. 1994). If the
Li depletion is related to the stellar age, one could expect a correlation
between lithium abundance and rotational velocity for stars of the same
mass, same metallicity and same spectral type (Skumanich 1972). In
agreement with this, Zahn (1992) and Pinsonneault et al. (1990) postulated
that the Li depletion in late-type stars is directly related to the loss of
angular momentum.\\

The age of clusters and consenquently that of their individual stars can be
determined from isochrone fitting. Cluster stars may therefore more efficiently
serve for
the study of changes in mixing-sensitive abundances and for the
search of anomalies in chemical compositions.  Open clusters
are moreover important laboratories, from the study of which one can address the question of Li evolution in
stellar interiors and surfaces, because they contain a significant sample of
stars in a wide range of mass with the same origin and same initial
chemical composition. In the last decades several observational diagnostics 
have discovered element abundances of stars belonging to open clusters of 
different ages with the help of high-resolution spectroscopy, and
even more recently with the multi-objects instruments.  
The evolution of Li and Be abundances
along specific parts of the colour-magnitude-diagrams (CMD) of open clusters
(main-sequence, turn-off (hereafter TO), subgiant and red giant branch) has been
investigated to test the physics and the extra-mixing processes included in stellar evolution models. 
Randich et al. (2002 \& 2007) have investigated both Be and Li abundances
in late-F and early-G main-sequence and subgiant stars from five different open
clusters with ages ranging from 50 Myr up to 4.5 Gyr. They confirmed shallow mixing 
in stellar interiors, which is able to transport surface material
deep enough for Li burning to occur, but not deep enough for Be
burning. Very recently, Smiljanic et al. (2010) have analysed  Be and Li abundances 
along the main sequence and RGB in the open cluster IC 4651 (with an age estimated around 1.5 Gyr). 
They show that both the evolution and dispersion of Be and Li abundances are successfully reproduced by  theoretical predictions from hydrodynamical models taking into account non-standard
mixing processes (rotation-induced mixing, atomic
diffusion, and thermohaline mixing).  \\

The present study focusses on evolved stars of the open cluster M67
(= NGC 2682), which has been used for the past three decades as an important
laboratory for studying stellar evolution (Burstein et al. 1986; Carraro
et al. 1996, and references therein). M67 is commonly mentioned as a solar
age cluster, i.e. one of the oldest open clusters of the Galaxy. However,
its age is still a matter of debate, and estimates vary from 3.5
to 4.5 Gyr.  A recent estimate (Sarajedini et al., 2009) based on the
comparison of a M67 proper-motion-cleaned colour-magnitude diagram to
theoretical isochrones points to low values around 3.5 - 4.0 Gyr,
confirming previous results from VandenBerg \& Stetson (2004) or Michaud et
al. (2004), who used evolutionary models taking into account microscopic
processes such as atomic diffusion.  On the other hand, spectroscopic studies
devoted to stars from M67 usually refer to a higher age of about 4.5
Gyr (e.g., Randich et al. 2006). Using spectroscopy and photometry, 
several authors determined an average iron abundance [Fe/H] for M67 
very close to the solar value.\\

Previous Li observations in M67 stars show that a real dispersion in the
lithium abundance of main-sequence objects exists even though 
at this phase the convective envelope is too superficial to reach the layer
where Li destruction can occur (Pasquini et al. 1997; Jones et al. 1999;
Randich et al. 2002; 2007). While standard models appeared inefficient to
explain this Li dispersion observed in unevolved M67 stars, the rotational
mixing seems to be a good candidate (Pasquini et al. 1997). Considering M67 subgiants, Sills \&
Deliyannis (2000) also find that stellar models with slow mixing caused by
rotation are the most suitable to explain the evolution of low-mass
stars. By measuring rotational velocities for 28 stars from the main-sequence 
to the giant branch in M67, Melo et al. (2001) provided an
analysis of the history of the angular momentum of stars of $1.2
M_\odot$. It verifies that these velocities probably obey different laws
for the evolution of the angular momentum on the main-sequence and on the
giant branch. \\

Our observations of M67 stars along an evolutionary sequence from the TO
to the top of the RGB offers the opportunity to shed light on
the interplay between Li and rotation evolution and to test the extra-mixing scenario in
stellar structures. In Section 2 we present the observational data and our
method to determine stellar parameters (effective temperature, surface
gravity, metallicity), Li abundances, and rotational velocities (\vsini).  In
Section 3 we present the stellar models we used to study the Li evolution
in evolved stars of M67, and our conclusions are drawn in Section 4.

\section{Spectroscopic observations and data analysis}

\begin{figure}[h]
   \centering
  \includegraphics[width=9cm]{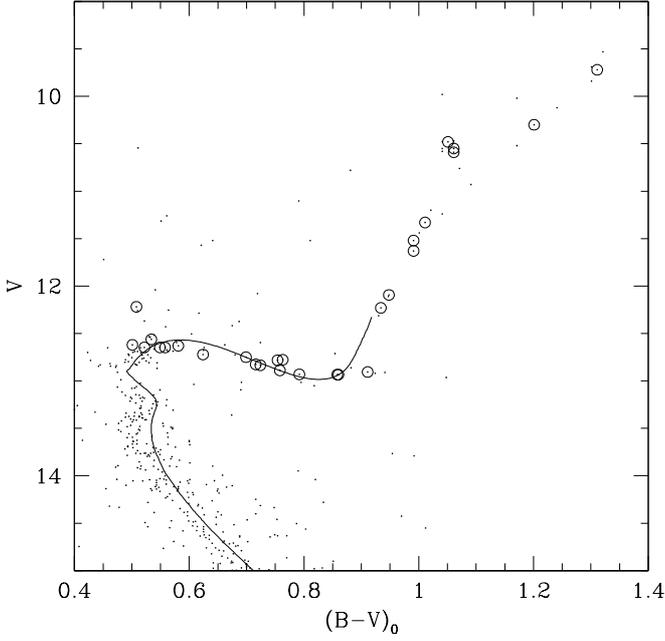}
     \caption{Colour-magnitude diagram of the open cluster M67.
    ($V$, $B-V$) photometry is taken from Montgomery et al. 
      (1993) and a colour excess of $E(B-V)=0.05$ has been assumed, as in Melo et al. (2001). 
      Stars from our sample are shown as open dot circles. The track
    is an isochrone at 3.7 Gyr from Vandenberg (private communication) based on 
    models with diffusion from Michaud et al. (2004).}
       \label{cm}
   \end{figure}

Our study is based on a sample of 28 post-main-sequence stars of
the open cluster M67. 
Figure \ref{cm} displays a colour-magnitude diagram of the open cluster
M67 (photometry from Montgomery et al., 1993). We overplotted a
best-fitting isochrone of 3.7 Gyr provided by VandenBerg (private
communication) based on the models with diffusion by Michaud et al. (2004).\\

These stars are divided into three groups regarding their evolutionary
stages: (i) TO stars, (ii) subgiant branch stars, and (iii)
RGB stars, including three clump
giants. According to a standard evolutionary scenario, TO and subgiant stars
must present a fast expansion of their convective envelopes, while giant
stars must have passed already through the first dredge-up episode,
bringing to the surface CNO products, hence modifying the surface
abundances. Let us mention that among our sample is the Li-rich subgiant S1242, which we analysed in detail
in Canto Martins et al. (\cite{Bruno06}).

\subsection{Spectroscopic observations}

The observations of these stars were carried out with the VLT/Unit2 ESO telescope (Paranal, Chile), using the FLAMES-UVES  spectrograph (Pasquini et al. 2002), during January 2004, January 2005 and March 2005. Using photometry and astrometry data of M67 from EIS pre-FLAMES (Momany et al. 2001) we selected our stars near the centre field with $\alpha=8:48$ and $\delta=+11:46$. We used two different fiber configurations. For the brightest stars ($V<12$, i.e., RGB stars) we observed four stars, with four fibers dedicated to the sky with the exposure time set to 1\,500 sec, providing a mean signal-to-noise ratio (S/N) of 50 (per pixel), over the lithium region (6\,700 - 6\,720 \AA). For the faintest stars ($V>12$, TO and subgiant stars), seven fibers were used for the stars and one for the sky. The exposure time for these groups was set to 1\,500, 2\,580, and 3\,000 sec, providing a S/N between 65 and 100 (per pixel). The stellar sample was observed with the UVES red arm centred at 580 nm, covering a wavelength range of 420-680 nm, and also centred at 860 nm, covering a larger wavelength range of 660-1\,060 nm. The observations have a resolving power of R $ \sim $ 47\,000 (1 arcsec sky aperture). Moreover, some stars (marked in Table \ref{log} with a star) we only observed with the UVES red arm centred at 580 nm.\\ 

The spectra were reduced with the standard FLAMES/UVES data reduction pipeline following the usual steps of reduction (bias, flat-field, and background corrections, order fiber definition, wavelength calibration of the spectra with a ThAr spectrum and extraction of the spectra).  Concerning the sky subtraction, for bright stars, four fibers were allocated for the sky on each plate. The average of the sky fibers was subtracted from each star spectrum. However, for the faintest stars only one fiber was dedicated to the sky and then subtracted from the star spectrum. Then we used IRAF to normalize them to a pseudo continuum and to bring the reduced spectra to rest. The radial velocities for our stellar sample were calculated with the IRAF {\tt fxcor} task. The stellar spectra were cross-correlated with a spectrum of the Sun (Hinkle et al. 2000). The shifts were then computed into radial velocities of the stars and heliocentric correction was applied. The results are presented in Table 1 and the errors in these velocities show a stability at a level of a few hundreds of m/s.\\

The probability that these stars belong to this stellar cluster is larger than 70\%, according to Sanders (1977) and their estimated radial velocities should be consistent with the estimated radial velocity for M67 of 33.0 km~s$^{-1}$ (Friel \& Janes 1993). This is not the case for the star S1000 with 0\% of membership probability and a radial velocity of  42.8 km~s$^{-1}$, which is very different from the group of stars and is therefore rejected.\\

Finally, we also combined different exposures collected at different dates  for some TO and subgiant stars,  in order to reach a good quality of the data and to increase the mean S/N over the Li region to 100 per pixel.\\

A complete log of observations is presented in Table \ref{log}.  
Star identifiers are from Sanders (1977). The radial velocity correction that was applied is mentioned in the last column (which agrees very well with the values produced by Melo et al. 2001).

\begin{table*}
\caption{Log of the spectroscopic observations of our 28 sample stars. The stars were observed using two different setups, except for those marked with ($\star$), which were only observed using the setup centred at 580 nm (see text).}
\centering
\begin{center}
\begin{tabular}{ccccccccc}
\noalign{\smallskip}
\hline \hline
\noalign{\smallskip}
Sanders	&	V	&	(B-V)	&	$\alpha_{2000}$ 	&	$\delta_{2000}$	&	Julian Day	&	Exposure	&	\vrad	\\
ID	&		&		&		&		&	+ 2453000	&	time (sec)	&	(\kms)	\\
\hline
\noalign{\smallskip}
S774	&	12.93	&	0.85	&	08:50:49.93	&	11:49:12.90	&	009.74	&	3000	&	33.5	\\
	&		&		&		&		&	011.77	&	1500	&	34.2	\\
	&		&		&		&		&	031.76	&	3000	&	34.0	\\
	&		&		&		&		&	032.76	&	3000	&	33.6	\\
	&		&		&		&		&	033.64	&	3000	&	34.0	\\
S806	&	12.78	&	0.81	&	08:51:00.17	&	11:54:32.10	&	009.74	&	3000	&	33.4	\\
	&		&		&		&		&	011.77	&	1500	&	34.3	\\
	&		&		&		&		&	031.76	&	3000	&	34.0	\\
	&		&		&		&		&	032.76	&	3000	&	33.6	\\
	&		&		&		&		&	033.64	&	3000	&	34.0	\\
S978	&	9.72	&	1.37	&	08:51:17.48	&	11:45:22.70	&	008.77	&	1500	&	34.7	\\
	&		&		&		&		&	008.79	&	1500	&	34.6	\\
S1000$^\star$	&	12.84	&	0.78	&	08:51:23.83	&	11:47:16.20	&	382.73	&	2580	&	42.8	\\
	&		&		&		&		&	382.76	&	2580	&	42.9	\\
S1010	&	10.48	&	1.11	&	08:51:22.81	&	11:48:01.80	&	008.77	&	1500	&	34.1	\\
	&		&		&		&		&	008.79	&	1500	&	33.8	\\
S1016	&	10.3	&	1.26	&	08:51:17.11	&	11:48:16.30	&	008.77	&	1500	&	34.6	\\
	&		&		&		&		&	008.79	&	1500	&	34.4	\\
S1034	&	12.65	&	0.61	&	08:51:18.55	&	11:49:21.70	&	028.74	&	3000	&	34.6	\\
	&		&		&		&		&	028.78	&	3000	&	34.6	\\
	&		&		&		&		&	033.69	&	3000	&	34.8	\\
	&		&		&		&		&	033.73	&	3000	&	34.9	\\
S1074	&	10.59	&	1.12	&	08:51:12.68	&	11:52:42.20	&	007.76	&	1500	&	34.2	\\
	&		&		&		&		&	007.78	&	1500	&	34.7	\\
S1231	&	12.93	&	0.92	&	08:51:29.38	&	11:45:27.80	&	009.74	&	3000	&	32.9	\\
	&		&		&		&		&	011.77	&	1500	&	33.6	\\
	&		&		&		&		&	031.76	&	3000	&	33.5	\\
	&		&		&		&		&	032.76	&	3000	&	33.1	\\
	&		&		&		&		&	033.64	&	3000	&	33.5	\\
S1239	&	12.75	&	0.76	&	08:51:44.03	&	11:46:24.60	&	028.74	&	3000	&	32.8	\\
	&		&		&		&		&	028.78	&	3000	&	32.9	\\
	&		&		&		&		&	033.69	&	3000	&	33.2	\\
	&		&		&		&		&	033.73	&	3000	&	33.2	\\
S1242	&	12.72	&	0.68	&	08:51:36.04	&	11:46:33.70	&	009.74	&	3000	&	37.5	\\
	&		&		&		&		&	011.77	&	1500	&	38.4	\\
	&		&		&		&		&	031.76	&	3000	&	34.1	\\
	&		&		&		&		&	032.76	&	3000	&	34.4	\\
	&		&		&		&		&	033.64	&	3000	&	35.1	\\
S1245	&	12.93	&	0.92	&	08:51:44.77	&	11:46:46.20	&	028.74	&	3000	&	33.1	\\
	&		&		&		&		&	028.78	&	3000	&	33.1	\\
	&		&		&		&		&	033.69	&	3000	&	33.3	\\
	&		&		&		&		&	033.73	&	3000	&	33.4	\\
S1254	&	11.52	&	1.05	&	08:51:45.08	&	11:47:46.00	&	007.76	&	1500	&	32.9	\\
	&		&		&		&		&	007.78	&	1500	&	33.5	\\
S1268	&	12.65	&	0.58	&	08:51:49.97	&	11:49:31.40	&	009.74	&	3000	&	32.7	\\
	&		&		&		&		&	011.77	&	1500	&	33.4	\\
	&		&		&		&		&	031.76	&	3000	&	33.2	\\
	&		&		&		&		&	032.76	&	3000	&	32.9	\\
	&		&		&		&		&	033.64	&	3000	&	33.3	\\
S1273	&	12.22	&	0.57	&	08:51:39.26	&	11:50:04.00	&	028.74	&	3000	&	34.6	\\
	&		&		&		&		&	028.78	&	3000	&	34.6	\\
	&		&		&		&		&	033.69	&	3000	&	34.7	\\
	&		&		&		&		&	033.73	&	3000	&	34.8	\\
S1275	&	12.56	&	0.59	&	08:51:37.43	&	11:50:05.40	&	009.74	&	3000	&	33.4	\\
	&		&		&		&		&	011.77	&	1500	&	34.1	\\
	&		&		&		&		&	031.76	&	3000	&	33.9	\\
	&		&		&		&		&	032.76	&	3000	&	33.7	\\
	&		&		&		&		&	033.64	&	3000	&	34.0	\\
S1277	&	11.63	&	1.05	&	08:51:42.35	&	11:50:07.80	&	008.77	&	1500	&	34.6	\\
	&		&		&		&		&	008.79	&	1500	&	34.3	\\
S1279	&	10.55	&	1.12	&	08:51:28.98	&	11:50:33.00	&	007.76	&	1500	&	33.3	\\
	&		&		&		&		&	007.78	&	1500	&	33.9	\\
S1288	&	11.33	&	1.07	&	08:51:42.37	&	11:51:23.10	&	007.76	&	1500	&	33.4	\\
	&		&		&		&		&	007.78	&	1500	&	34.0	\\
S1293	&	12.09	&	1.01	&	08:51:39.42	&	11:51:45.90	&	028.74	&	3000	&	34.2	\\
	&		&		&		&		&	028.78	&	3000	&	34.2	\\
	&		&		&		&		&	033.69	&	3000	&	34.3	\\
	&		&		&		&		&	033.73	&	3000	&	34.5	\\
\hline\hline
\end{tabular}
\label{log}
   \end{center}
\end{table*}

\begin{table*}
\begin{flushleft}
{\bf Table 1.} Cont.
\end{flushleft}
\centering
\begin{center}
\begin{tabular}{ccccccccc}
\noalign{\smallskip}
\hline \hline
\noalign{\smallskip}
Sanders	&	V	&	(B-V)	&	$\alpha_{2000}$ 	&	$\delta_{2000}$	&	Julian Day	&	Exposure	&	\vrad	\\
ID	&		&		&		&		&	+ 2453000	&	time (sec)	&	(\kms)	\\
\hline
\noalign{\smallskip}
S1305$^\star$	&	12.23	&	0.99	&	08:51:35.79	&	11:53:35.00	&	398.79	&	2580	&	34.3	\\
	&		&		&		&		&	431.68	&	2580	&	34.2	\\
S1319$^\star$	&	12.91	&	0.97	&	08:51:48.80	&	11:56:52.0	&	398.79	&	2580	&	34.4	\\
	&		&		&		&		&	431.68	&	2580	&	34.3	\\
S1323	&	12.83	&	0.78	&	08:51:35.43	&	11:57:56.80	&	028.74	&	3000	&	33.2	\\
	&		&		&		&		&	028.78	&	3000	&	33.2	\\
	&		&		&		&		&	033.69	&	3000	&	33.5	\\
	&		&		&		&		&	033.73	&	3000	&	33.5	\\
S1438$^\star$	&	12.89	&	0.82	&	08:52:11.35	&	11:45:37.50	&	398.79	&	2580	&	33.1	\\
	&		&		&		&		&	431.68	&	2580	&	33.0	\\
S1487$^\star$	&	12.63	&	0.64	&	08:52:04.81	&	11:58:28.80	&	398.79	&	2580	&	33.4	\\
	&		&		&		&		&	431.68	&	2580	&	33.3	\\
S1607$^\star$	&	12.62	&	0.56	&	08:52:21.45	&	11:50:41.70	&	398.79	&	2580	&	33.7	\\
	&		&		&		&		&	431.68	&	2580	&	33.7	\\
S2207$^\star$	&	12.65	&	0.62	&	08:51:32.47	&	11:47:52.50	&	398.79	&	2580	&	31.9	\\
	&		&		&		&		&	431.68	&	2580	&	31.9	\\
S2208	&	12.78	&	0.82	&	08:51:32.42	&	11:48:01.30	&	028.74	&	3000	&	33.0	\\
	&		&		&		&		&	028.78	&	3000	&	33.0	\\
	&		&		&		&		&	033.69	&	3000	&	33.1	\\
	&		&		&		&		&	033.73	&	3000	&	33.1	\\
\hline\hline
\end{tabular}
   \end{center}
\end{table*}

\subsection{Spectral synthesis}

We performed spectral synthesis analysis on our 27 sample stars, to derive 
stellar parameters (\Teff, \logg, $\xi$, [Fe/H]) and to
measure rotational velocities and lithium abundances. We used the 
MARCS models of stellar atmospheres (Gustafsson et
al. 2008), which are based on plan-parallel and spherical models at local
thermodynamical equilibrium (LTE).  The turbospectrum spectral synthesis
tools (Alvarez \& Plez 1998) and interpolation routines on model
atmospheres were used. Solar abundances were taken from Asplund,
Grevesse \& Sauval (2005) and the collisional damping treatment was performed 
based on the work of Barklem et al. (Barklem et al. 2000a,b; Barklem \& Piskunov 2003, 
Barklem \& Aspelund-Johansson 2005).  To compute synthetic spectra, we took 
taken into account atomic (see below) and molecular line lists: TiO (Plez 1998), VO (Alvarez \&
Plez, 1998), and CN and CH (Hill et al. 2002). \\

In order to improve the precision on the atmospheric parameters and on the
 Li abundances (\Ali~= $log \left(\frac{n_{Li}}{n_H}\right)+12$) presented
 in Canto Martins et al. (2007), we analysed our sample using the curve-of-growth 
method, which is based on the measurement of equivalent widths (EWs) of  a large number of iron lines
 present in our FLAMES-UVES observations. Indeed, weak 
 features (mainly caused by the Fe I line at 6\,707.44 \AA) are known to blend
 the Li line (at 6\,707.78 \AA), which is used to derive \Ali . Hence, a very good
 precision on the estimate of the Fe abundance and on all atmospheric parameters
 is required to ensure a determination of
 \ALi with a high accuracy. \\

To do so, we first calibrated the oscillator strength values ($\log gf$) of 92 Fe I and 14 Fe II spectral
 lines issued from the VALD database (Kupka et al. 1999). These lines appear reasonably
unblended throughout the spectral domain (4\,200-8\,000~\AA) of the high-resolution spectra 
of the Sun and of Arcturus (Hinkle et al. 2000). For each line
the central wavelength, excitation potential ($\chi_{exc}$), and our value on
$\log gf$ improved by an inverse solar and Arcturus analysis are presented in Table~\ref{Felines} (online material). 
The equivalent widths (EW)  from the Sun and from Arcturus spectra (Hinkle et al. 2000) are also reported. They were  
measured with an automatic procedure using the DAOSPEC package (Stetson \& Pancino 2008). 
In order to check our corrections on $\log gf$ values, we used our measured EWs to derive the FeI and
FeII abundances for the Sun and for Arcturus. For the Sun, we adopted the
following atmospheric parameters: \Teff = 5777 K (Neckel 1986), $\xi$ = 1.0
km s$^{-1}$ (Ruedi et al. 1997), $\log g$ = 4.44 (Allen 1973); and for 
Arcturus: \Teff~=4300 K, $\xi$ = 1.6 km s$^{-1}$ , $\log g$ =
1.8 (Zoccali et al. 2004). The solar iron abundances we derived, $\log n(Fe I) = 7.49 \pm 0.03$ dex and $\log n(Fe II) =
7.48 \pm 0.04$ dex, agree well with the values of Asplund, Grevesse
\& Sauval (2005) and also with the value of the Fe II abundance found by
Bi\'emont et al. (1991).  For Arcturus we found $\log n(Fe I) = 7.02 \pm
0.05$ dex and $\log n(Fe II) = 7.05 \pm 0.06$ dex, which agrees with
determinations from Peterson et al. (1993) and from Carraro et
al. (2004).\\

\begin{figure}[hbp]
  \centering
   \includegraphics[width=9cm]{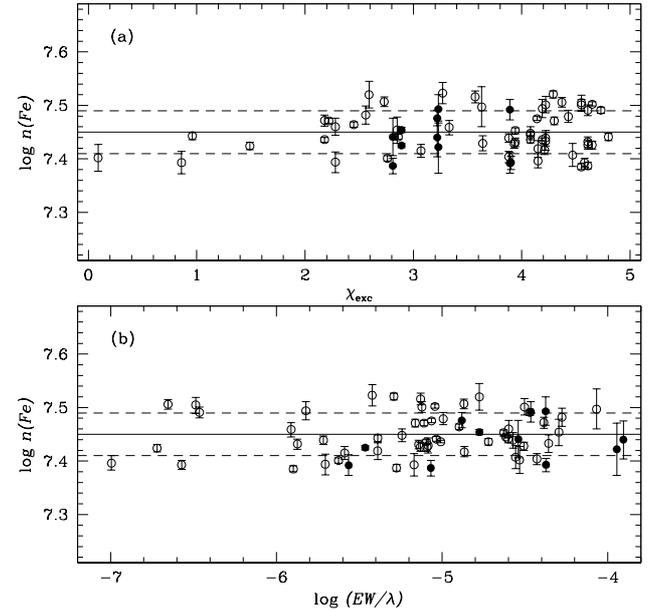}
   \caption{Analysis of the turn-off star S1275 : {\bf (a)} Fe lines excitation equilibrium. {\bf (b)} relations between Fe abundances and equivalent widths. 
   		Open and filled circles represent the Fe I and Fe II lines, respectively. The solid line represents the mean iron abundance and the dashed lines represent 
		the 1$\sigma$ of the distribution.}
\label{parameters}
\end{figure}

Then we estimated for our 27 sample stars the effective temperatures, microturbulent velocities, and stellar surface gravities by imposing an excitation equilibrium for the Fe I abundances, an equilibrium between the Fe I abundances and the equivalent widths, and a FeI/FeII ionization equilibrium as presented in Fig. \ref{parameters} for the TO star S1275. The initial values adopted for the atmospheric parameters were the ones published in Canto Martins et al. (2007). In this previous work, the \Teff were estimated from photometry (Montgomery et al. 1993; Houdashelt et al. 2000), from the H$_\alpha$ and H$_\beta$ lines and also the Fe I lines ratio in the Li region. The \logg was estimated using evolutionary models. For the metallicities, the solar metallicity was adopted as a first guess and then adjusted by fitting the Fe I lines for the Li region.\\

With this method the metallicities [Fe/H]  were calculated for a solar 
abundance $\log n(Fe)_\odot = 7.49$  (as derived here) and the [Fe/H] errors are based on the standard
deviations in the Fe I lines abundances only. The accuracies obtained on
atmospheric parameters are $\pm$70 K for \Teff, $\pm$0.2 dex for $\log g$
and $\pm$0.2~km~s$^{-1}$ for $\xi$.\\

Table \ref{atm_estrelas_M67} provides the atmospheric parameters adopted to
compute the synthetic spectra used to derive the projected rotational velocity
(\vsini) and \Ali for each star in our sample. To derive the projected rotational velocity, we used the same procedure as in de
Medeiros et al. (2006). We convolved the resulting spectra  (taking into account the
instrumental profile of FLAMES-UVES)  with a
rotational profile to adjust the broadening observed on a group of
FeI lines in the Li region (6\,700--6\,720 \AA).  The \vsini~ values we
measured (presented in Table \ref{atm_estrelas_M67}) have an accuracy of $\pm$1.0 km s$^{-1}$.\\

\begin{figure}[!ht]
   \centering
   \includegraphics[width=9cm]{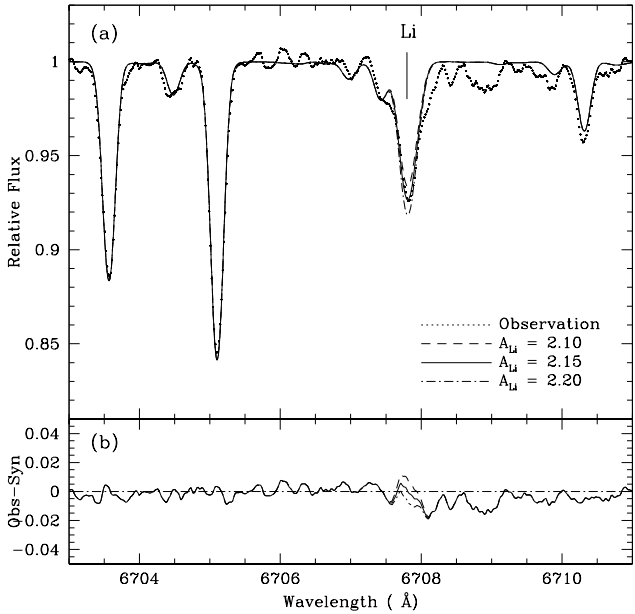}
      \caption{Lithium line spectroscopic region of the turn-off star S1275:  {\bf  (a)} observation (dotted line) and synthetic spectra computed with the atmospheric parameters listed in Table \ref{atm_estrelas_M67}, and with \Ali = 2.10, 2.15, 2.20 dex (dashed, solid and dashed-dot lines respectively). The best synthetic spectrum is clearly the one computed with \Ali = 2.15. {\bf (b)} Residual difference between synthetic and observed spectra. }
       \label{li_s1275}
   \end{figure}

Finally, the determination of \Ali was made by the best fit between
observed and synthetic spectra, obtained on the Li line at 6707.78 \AA. For
each star, the total error on the Li abundance was estimated by
computing the quadratic sum of errors induced by errors on individual
parameters (\Teff, \logg, $\xi$, [Fe/H], \vsini) and also errors associated with the best-fit determinations.
As discussed by  De Laverny et al. (2003) and L\`ebre et al. (2006), the main source of uncertainty in the 
abundances determinations is caused by errors in the \Teff determination. They have estimated that \Teff 
measurements with errors smaller than 200\,K lead to errors smaller than 0.2\,dex in the derived 
metallicity and lithium abundance. We made some tests, and the obtained errors are about the same 
as proposed in these previous studies. However, we found that errors in
the parameters $\log g$ and $\xi $ also influence the determination of the
metallicity of the stars in our sample. We found that the
errors obtained in [Fe/H] caused by errors in $\log g$ are smaller than those
measured when we analyse the errors in the metallicity caused by $\xi$.
Taking into account the two sources of errors, we obtained a maximum error
of 0.16 dex in [Fe/H]. In its last column, Table \ref{atm_estrelas_M67} also presents \Ali values and their
accuracies. As an illustration we show in Fig. \ref{li_s1275} a zoom on the Li region together with 
synthetic spectra for several values of \Ali for the TO star S1275.\\

The stellar parameters, microturbulent and rotational velocities and \Ali measured in this study are more 
accurate than the values produced by Canto Martins et al. (2007) because of the large number of Fe I and Fe II 
lines used in the characterization of the stellar sample.

\begin{table}[!h]
\caption{Estimated atmospheric parameters for our  sample stars. Rotational velocities and lithium abundances 
derived from  our spectral synthesis analysis are also given. The identifiers (ID) are from Sanders (1977).} 
\label{atm_estrelas_M67}
\begin{center}
\begin{tabular}{@{}ccc@{}c@{}c@{}c@{}@{}c@{}}
\hline\hline
ID & $T_{eff}$&	$\log g$&      $\xi$&     $[Fe/H]$	&$v \sin i$&$A_{Li}$	\\
   & (K)      &	(dex)        &      (\kms)&  (dex)   	&(\kms)&	\\
\hline 
\noalign{\smallskip}
\multicolumn{7}{c}{Turn-off stars}\\
\hline

\noalign{\smallskip}
S1273	&	6159	&	4.01	&	1.86	&	$-$0.04$\pm$0.04	&	8.0	&	1.90$\pm$0.11	\\
S1607	&	6127	&	3.81	&	1.77	&	$-$0.11$\pm$0.06	&	3.5	&	1.70$\pm$0.10	\\
S1275	&	6050	&	4.00	&	1.64	&	$-$0.04$\pm$0.04	&	4.9	&	2.15$\pm$0.11	\\
S1034	&	6020	&	3.90	&	1.94	&	$-$0.08$\pm$0.03	&	4.0	&	1.30$\pm$0.11	\\
S2207	&	6000	&	3.90	&	1.62	&	$-$0.03$\pm$0.03	&	4.0	&	1.20$\pm$0.12	\\
S1268	&	5996	&	3.86	&	1.75	&	$-$0.11$\pm$0.07	&	3.5	&	0.90$\pm$0.12	\\
S1487	&	5940	&	3.81	&	1.67	&	$-$0.05$\pm$0.03	&	2.5	&	1.10$\pm$0.13	\\
\hline
\noalign{\smallskip}
\multicolumn{7}{c}{Subgiant stars}\\
\hline
\noalign{\smallskip}
S1242	&	5810	&	3.90	&	1.50	&	$-$0.04$\pm$0.06	&	6.1	&	2.70$\pm$0.13	\\
S1323	&	5654	&	3.90	&	1.50	&	$+$0.03$\pm$0.04	&	3.3	&	$<$1.00		\\
S1239	&	5644	&	3.80	&	1.41	&	$+$0.00$\pm$0.03	&	3.1	&	$<$0.80		\\
S806	&	5461	&	3.80	&	1.25	&	$+$0.08$\pm$0.03	&	4.1	&	0.00$\pm$0.19	\\
S2208	&	5429	&	3.90	&	1.24	&	$+$0.01$\pm$0.03	&	3.6	&	0.70$\pm$0.14	\\
S1438	&	5420	&	3.80	&	1.27	&	$-$0.06$\pm$0.03	&	2.2	&	$<$0.00		\\
S774	&	5240	&	3.70	&	1.20	&	$-$0.08$\pm$0.03	&	3.2	&	$<$0.00		\\
\hline
\noalign{\smallskip}
\multicolumn{7}{c}{Giant stars}\\
\hline
\noalign{\smallskip}
S1245	&	5170	&	3.61	&	1.19	&	$-$0.02$\pm$0.03	&	3.0	&	0.00$\pm$0.15	\\
S1231	&	5130	&	3.60	&	1.12	&	$-$0.02$\pm$0.03	&	3.0	&	$<$0.10		\\
S1319	&	5104	&	3.61	&	1.16	&	$-$0.07$\pm$0.03	&	3.0	&	$<$0.40		\\
S1293	&	4970	&	3.30	&	1.32	&	$-$0.01$\pm$0.03	&	3.2	&	$-$0.20$\pm$0.22\\
S1305	&	4940	&	3.20	&	1.18	&	$-$0.08$\pm$0.03	&	2.8	&	$<-$0.20		\\
S1254	&	4820	&	2.91	&	1.30	&	$-$0.03$\pm$0.04	&	3.0	&	$<-$0.40		\\
S1277	&	4820	&	3.00	&	1.26	&	$+$0.01$\pm$0.05	&	3.0	&	$<-$0.40		\\
S1279	&	4779	&	2.72	&	1.57	&	$-$0.01$\pm$0.08	&	2.5	&	$-$0.50$\pm$0.23\\
S1288	&	4773	&	2.90	&	1.32	&	$-$0.01$\pm$0.04	&	2.5	&	$-$0.20$\pm$0.21\\
S1074	&	4750	&	2.60	&	1.62	&	$-$0.07$\pm$0.04	&	2.5	&	$-$0.40$\pm$0.24\\
S1010	&	4748	&	2.60	&	1.58	&	$-$0.03$\pm$0.07	&	2.5	&	$<$0.00		\\
S1016	&	4430	&	2.31	&	1.52	&	$-$0.05$\pm$0.05	&	2.0	&	$-$0.50$\pm$0.25\\
S978	&	4260	&	1.90	&	1.63	&	$-$0.15$\pm$0.06	&	2.0	&	$-$1.00$\pm$0.27\\
\hline\hline
\end{tabular}
\end{center}
\end{table}

\begin{table}[h]
\caption{Comparison for the star S1010 between atmospheric parameters from Tautvai\v{s}iene et al. (2000) (T00), Young et al. (2006) (Y05), and Pancino et al. (2010) (P10) and those derived in the present work.} 
\label{metal}
\begin{center}
\begin{tabular}{c|c|c|c|c}
\hline \hline
Parameters	&	T00 &	Y05	&	P10 &	Our work	\\
\hline	
\Teff (K)	&	4730$\pm$100&	4700$\pm$100	&	4650$\pm$100	&4748$\pm$70	\\
\logg	&	2.4$\pm$0.3&	2.3$\pm$0.3	&	2.8$\pm$0.2	&2.6$\pm$0.20	\\
$\xi$ (km s$^{-1}$)	&	1.6$\pm$0.3	&1.34$\pm$0.2	&	1.3$\pm$0.1	&1.58$\pm$0.20	\\
$[Fe/H]$	&	-0.01$\pm$0.11	&0.00$\pm$0.10	&	0.06$\pm$0.01&	-0.03$\pm$0.07	\\
\hline \hline					
\end{tabular}
\end{center}
\end{table}

\begin{table}[!h]
\caption{Comparison for S1034 and S1239 between atmospheric parameters from Randich et al. (2006) and lithium abundances from Randich et al. (2007) and those derived in the present work.} 
\label{comparison}
\begin{center}
\begin{tabular}{c|c|c|c|c}
\hline \hline
	&	\multicolumn{2}{|c|}{S1034}	&	\multicolumn{2}{|c}{S1239}	\\
\cline{2-5}
	&	Randich &	Our work	&	Randich &	Our work	\\
\hline	
\Teff (K)	&	5969$\pm$70&	6020$\pm$70	&	5477$\pm$70	&5644$\pm$70	\\
\logg	&	4.0$\pm$0.25&	3.9$\pm$0.20	&	3.8$\pm$0.25	&3.8$\pm$0.20	\\
$\xi$ (km s$^{-1}$)	&	1.5$\pm$0.15	&1.94$\pm$0.20	&	1.25$\pm$0.15	&1.41$\pm$0.20	\\
$[Fe/H]$	&	$+$0.01$\pm$0.05	&-0.08$\pm$0.03	&	$+$0.02$\pm$0.03&	$+$0.00$\pm$0.03	\\
\Ali	&	$<$1.14&	1.30$\pm$0.11	&	1.12$\pm$0.10	&$<$0.80	\\
\hline	\hline					
\end{tabular}
\end{center}
\end{table}

\subsection{Comparison with other spectroscopic studies}

We compared the parameters we determined with previous spectroscopic
studies available in the literature. Concerning the metallicity of
M67 we found from our sample stars a mean metallicity for our evolved
stars of $[Fe/H]=-0.05\pm0.04$. This value agrees very well with
the lowest values from the literature as $[Fe/H]=-0.05$ (Canterna et al. 1986), 
$[Fe/H]=-0.07$ (Anthony-Twarog 1987), $[Fe/H]=-0.08$ (Friel \& Jane 1991), 
$[Fe/H]=-0.09$ (Friel \& Jane 1993), $[Fe/H]=-0.05$ (Shetrone \& Sandquist 2000) and $[Fe/H]=-0.03$ (Tautvai\v{s}iene et al. 2000). 
However, other studies point to higher values for the mean metallicity of M67 as $[Fe/H]=0.06$ (Nissen et al. 1987), 
$[Fe/H]=0.04$ (Garcias Lopez et al. 1988), $[Fe/H]=0.02$ (Friel \& Boesgaard 1992), and recently, $[Fe/H]=0.02\pm0.14$ (Yong et al. 2005), 
$[Fe/H]=0.03\pm0.01$ (Randich et al. 2006, hereafter R06) and $[Fe/H]=0.05\pm0.02$ (Pancino et al. 2010). \\

Comparing some stars in our sample with other studies in greater detail, we have the star S1010 that was also studied by Tautvai\v{s}iene et al. (2000), Yong et al. (2005), and Pancino et al. (2010). Table \ref{metal}  compares the values of the atmospheric parameters and metallicity measured for this star. Evidently the atmospheric parameters obtained in all studies agree well. However, we measured a lower metallicity compared to their results. One possible explanation for this is the higher solar iron abundance adopted by these studies ($\log n(Fe I) = 7.54$ for Yong et al. 2006 and $\log n(Fe I) = 7.50$ for Pancino et al. 2010). Another explanation is that both the superficial gravity and the microturbulent velocity affect the metallicity
of the star. In this case, as the value of $\xi$ obtained in our study is
higher than that obtained by Yong et al. (2005) and Pancini et al. (2010),
this difference leads to errors on [Fe/H] smaller than 0.12 dex in S1010. This difference may also explain the low mean metallicity of M67 found in our study.\\

\begin{figure}[h]
   \centering
   \includegraphics[width=9cm]{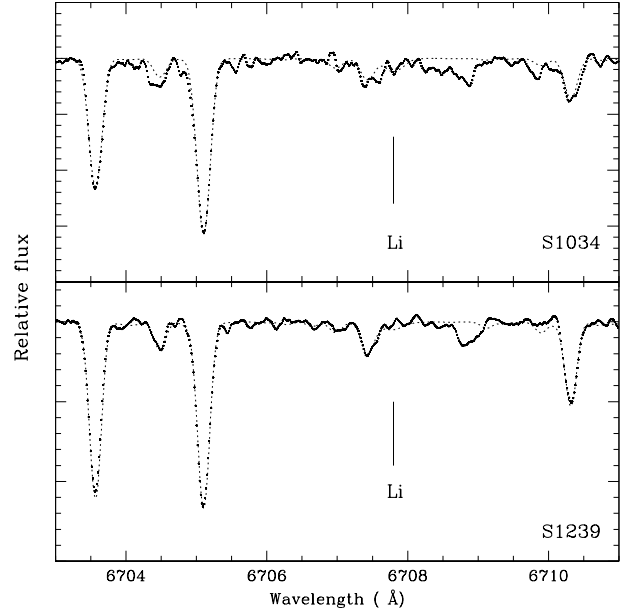}
      \caption{Li region of the stars S1034 and S1239, which were also studied by Randich et al. (2006). Observations are represented with a dotted line and synthetic spectra by a dashed line. The vertical line indicates the Li feature at 6707.78 \AA.}
       \label{s1034_s1239}
   \end{figure}

Two other targets,
S1034 and S1239 (Fig. \ref{s1034_s1239}), were also included in the spectroscopic study of Randich et al. (2006). For these
two objects, Table \ref{comparison} provides the comparison of the
stellar parameters between R06 and the present study. Both analyses appear
to agree fairly well within the error bars for stellar parameters determination, even if we derive 
a lower metallicity for S1034 (but still consistent within the error bars) and a higher microturbulent velocity. 
However, the \Ali values present differences that may be explained by the adopted methodology 
to compute the model atmosphere and synthetic spectra.\\

Finally, our rotational velocities (\vsini) agree well with the results obtained by Melo
et al. (2001) on evolved stars of M67. Moreover our study statistically
extends the number of M67 evolved stars with measured \vsini.  In this
way, our efforts complement the existing values of atmospheric parameters
and \vsini~ values for M67 objects, and can assist studies that are mainly dedicated
to the evolution of the angular momentum in the open cluster M67.\\

Up to date, as mentioned in different studies (Pilachowski et al. 1988,
Balachandran 1995), subgiant and giant stars in M67 show no significant
\Ali, which points to a severe Li depletion after the TO. For unevolved stars of 
M67, Jones et al. (1999) and Randich et
al. (2002, 2007) show a large dispersion on the \Ali. Balachandran (1995)
also studied the Li abundance in M67 evolved stars, but only upper limits were
estimated. In this context, the analysis we performed on our sample
complements the previous observational studies mainly devoted to stars of
M67 near the TO. Our data present the same dispersion as established by Jones et al.  (1999) 
and Randich et al. (2002 and 2007) and shows a gradual decrease of \ALi for stars  cooler than 5500 K (see 
Fig. \ref{li_teff}, next section). The present result points to a
depletion of Li in excess to standard predictions as pointed out
by Pasquini et al. (1997).

\section{Connecting the Li abundances and the rotational velocities}

The lithium abundances derived for our stars clearly indicate that this
element has been depleted at their surface already during the
main sequence evolution. In an attempt to consistently explain the
evolution of both the lithium abundances and the rotational velocities, 
we computed a grid of stellar evolution models with rotation and compared the
predictions to our sample.

\subsection{Stellar evolution models with rotation}

We computed a grid of stellar evolution models in the mass range 1.30
M$_\odot$ to 1.37 M$_\odot$ (see Table~\ref{tab:models}) including 
rotation,
microscopic diffusion, and thermohaline mixing (as described in Charbonnel
\& Zahn \cite{CCJPZ07}) using
STAREVOL V3.00 (Siess \cite{Siess06},; Siess et al. \cite{Siess08};
Decressin et al. \cite{Decressin09a}). We used the solar abundances given by 
Asplund et al. (\cite{AGS05}) as a reference for
our stellar evolution models, and the corresponding opacities were
generated with the OPAL Opacity code (Iglesias \& Rogers 
\cite{OPAL96})\footnote{The tables were generated using the web interface available at 
http://opalopacity.llnl.gov/}. Convective zones are defined
using the Schwarzschild criterion and the mixing length formalism with an
adopted mixing length parameter $\alpha_{\rm MLT} = 1.78$ that was
calibrated on the Sun. The models have a metallicity of [Fe/H] = -0.03,
which translates into a metal content Z = 0.01148. The initial helium mass
fraction Y$_{\rm ini}$ = 0.2688 was calibrated for the solar
model. The initial lithium content is set to \Ali = 3.20\footnote{This
  abundance is obtained using the solar meteoritic abundance given by
   Aslpund et al. (\cite{AGS05}) scaled down to
the metallicity of M67.}. Mass loss is included in our models from the
ZAMS
and beyond following Reimers (\cite{Reimers75}). Microscopic diffusion 
(gravitational settling) of
chemical species is accounted for using the
  approximation of Paquette et al. (\cite{Paquette86}) for the 
microscopic diffusion
  coefficient and the expressions given by Montmerle \& Michaud 
(\cite{MM76}) for the
  microscopic diffusion velocity.
We adopt the Maeder \& Zahn (\cite{MZ98}) formalism to account for the
transport of angular momentum and of chemical species by meridional
circulation and shear-induced turbulence in the models. The formalism and
its introduction in STAREVOL are described at length in
Palacios et al. (\cite{Palacios03,PCTS06}), and we refer the reader to these papers for
further details. Let us nonetheless recall that we use
the prescription from Zahn (\cite{Zahn92}) for the horizontal turbulent 
diffusion
coefficient $D_h$ with $C_h = 1.0$.  We also assume that convective
regions
rotate as solid bodies. For those models with equatorial
velocity at the ZAMS of about 19 km.s$^{-1}$, we assume angular 
momentum
losses associated with a magnetic torque as described by
Kawaler (\cite{Kawaler88}):
\begin{equation}
\left(\frac{dJ}{dt}\right) \, = \, -K 
\Omega3\left(\frac{R}{R_\odot}\right)^{1/2}\left(\frac{M}{M_\odot}\right)^{-1/2}~~~~~(\Omega< \Omega_{sat})
\end{equation}
\begin{equation}
\left(\frac{dJ}{dt}\right) \, = \, -K \Omega 
\Omega_{sat}2\left(\frac{R}{R_\odot}\right)^{1/2} 
\left(\frac{M}{M_\odot}\right)^{-1/2}~~~~~(\Omega \geq \Omega_{sat}).
\end{equation}
 The parameter $K$ in Kawaler's law is related to the magnitude of the
   magnetic field strength, and $\Omega_{sat} = 14 \Omega_\odot$ following
   Bouvier et al. (\cite{Bouvier97}).
Lacking observational data on the ZAMS to constrain
the evolution of angular momentum of our sample stars, we use the v$\sin
i$
derived by Melo et al. (\cite{Melo01}) for main-sequence stars of M67 as a
guideline to calibrate the braking. For masses 1.30 to 1.37 M$_\odot$ we
adopt K = 8$\times$10$^{29}$.\\

We checked that a model of 1.25 M$_\odot$ including microscopic
diffusion, no rotation, and a mixing length parameter calibrated for the
Sun
reaches the turn-off at 3.75 Gyr, in fair agreement with the value
obtained
by Michaud et al. (\cite{Michaud04}), who used a more detailed treatment for
microscopic diffusion and radiative forces.
Bellow we will therefore consider
an age of 3.7 Gyr for M67 and all discussed predictions will refer
to values obtained when the models reach this age.\\
\begin{figure}[ht]
   \centering
   \includegraphics[width=9cm]{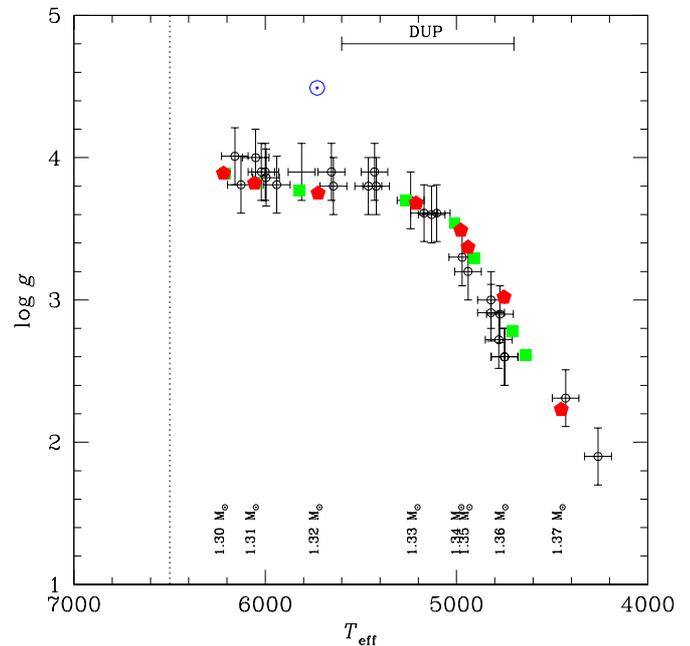}
      \caption{\logg versus \Teff~for our sample of stars, represented 
by black
      circles. Values at 3.7 Gyr for the rotating models with and without
      angular momentum extraction are represented by filled red pentagons
      and filled green squares respectively. The dotted line indicates the
      turn-off (TO) position and  the location of the first DUP is
      also indicated.}
       \label{teff_logg}
   \end{figure}

Table~\ref{tab:models} presents the results of our modelling in terms of
effective temperature and gravity, luminosity, equatorial velocity, and
lithium abundance at the ZAMS, the turn-off (e.g. TO) and 3.7 Gyr (e.g. 
M67). Models with and without
braking have an initial angular momentum of about 5.2$\times$10$^{49}$ 
and 3.2$\times$10$^{49}$ g.cm$^2$.s$^{-1}$ respectively.

\begin{table*}
\caption{Characteristics of stellar evolution models with
 rotation. Effective temperature, surface gravity, luminosity, surface
 rotation velocity, lithium abundance, age, and associated evolutionary
 phase are given for models with initial masses between 1.30 M$_\odot$ and
 1.37 M$_\odot$. In the last column, ``M67'' refers
 to the models predictions at the adopted age for the cluster of 3.7
 Gyr. For each quantity, the values for models without ($K_{\rm K} = 0$)
 and with ($K_{\rm K} \neq 0$) magnetic braking on the ZAMS are indicated,
 the latter is shown in bold face.}
\label{tab:models}
\begin{center}
\begin{tabular}{ccc|cc|cc|cc|cc|cc|c}
\hline\hline
mass    &  \multicolumn{2}{c}{T$_{\rm eff}$} & \multicolumn{2}{c}{log g} & \multicolumn{2}{c}{L} & \multicolumn{2}{c}{v$_{\rm surf}$} &\multicolumn{2}{c}{\ALi} & \multicolumn{2}{c}{age} & phase    \\
(M$_\odot$) & \multicolumn{2}{c}{(K)} & \multicolumn{2}{c}{(dex)} &
   \multicolumn{2}{c}{(L$_\odot$)} & \multicolumn{2}{c}{(km/s)} &
   \multicolumn{2}{c}{(dex)} & \multicolumn{2}{c}{(Gyr)} & \\
\hline
& $K_{\rm K} = 0$ & $K_{\rm K} \neq 0$ & $K_{\rm K} = 0$ & $K_{\rm K} \neq 0$ & $K_{\rm K} = 0$ & $K_{\rm K} \neq 0$ & $K_{\rm K} = 0$ & $K_{\rm K} \neq 0$ & $K_{\rm K} = 0$ & $K_{\rm K} \neq 0$ & $K_{\rm K} = 0$ & $K_{\rm K} \neq 0$ & \\
\hline
1.30 & 6699 &  6701 & 4.37 &  4.37 & 2.75 &  2.75 & 12.05 &
 18.9& 3.18 &
  3.18 & 4.48 10$^{-2}$ &   4.76 10$^{-2}$ & ZAMS\\
    & 6439 &    6440 & 3.99 &   3.99 & 5.55 &   5.57 & 10.86
&   8.88 & 2.99 &   1.26 & 3.36 &   3.38 & TO\\
    & 6212 &  6218 & 3.89 &   3.89 & 6.11 &   6.15 & 7.97 &
  8.03 & 2.93 &   1.14 & 3.7 &   3.7 & M67\\
\hline
1.31 & 6729 &   6729 & 4.36 &   4.36 & 2.85 &   2.85 & 11.92 &
  18.87 & 3.18 &   3.18 & 4.50 10$^{-2}$ &   4.50 10$^{-2}$ & ZAMS\\
    & 6457 &    6472& 3.99 &   3.99 & 5.70 &   5.70 & 10.72
&   9.00 & 2.93 &   1.27 & 3.24 &   3.24 & TO\\
    & 6050 &  6056 & 3.82 &   3.82 & 6.48 &   6.49 & 7.40 &
  7.22 & 2.80 &   1.08 & 3.7 &   3.7 & M67\\
\hline
1.32 & 6757 &   6758 & 4.36 &   4.36 & 2.95 &   2.96 & 11.88 &
  18.90 & 3.18 &   3.18 & 4.26 10$^{-2}$ &   4.26 10$^{-2}$ &ZAMS\\
    & 6476 &   6499 & 3.99 &   4.00 & 5.88 &   5.79 & 10.75 &
  9.05 & 2.93 &   1.23 & 3.19 &   3.11 &TO\\
    & 5824 &   5724 & 3.77 &   3.75 & 6.38 &   6.20 & 5.84 &
  5.73 & 2.65 &   0.93 & 3.7 &  3.7 & M67\\
\hline
1.33 & 6787 &   6787 & 4.36 &   4.36 & 3.06 &   3.06 & 12.09 &
  18.92 & 3.17 &   3.18 & 3.88 10$^{-2}$ &   4.02 10$^{-2}$ &ZAMS\\
    & 6479 &   6508 & 3.97 &   3.99 & 6.11 &   6.05 & 10.71 &
  9.15 & 2.92 &   1.32 & 3.14 &   3.07 &TO\\
    & 5266 &   5211 & 3.70 &   3.68 & 4.98 &   4.89 & 3.63 &
  4.14 & 1.82 &   0.19 & 3.7 &   3.7 & M67\\
\hline
1.34 & 6814 &   6808 & 4.35 &   4.35 & 3.16 &   3.17 & 12.09 &
  18.97 &  3.17 &   3.17 &  3.66 10$^{-2}$ &   3.78 10$^{-2}$ &ZAMS\\
    & 6502  &  6526 & 3.97 &   3.98 & 6.28 &   6.22  & 10.71
&   9.24 & 2.93 &   1.37& 3.05 &   2.99 & TO\\
    & 5010 &   4978 & 3.54 &   3.49 & 5.88 &   6.59 & 2.54 &
  3.08 &  1.44 &   -0.15 & 3.7 &  3.7 & M67\\
\hline
1.35 & 6841 &   6842 & 4.35 &   4.35 & 3.28  &   3.28 & 12.09 &
  18.97 & 3.17 &   3.18 & 3.38 10$^{-2}$ &   3.56 10$^{-2}$ & ZAMS\\
    & 6552 &   6544 & 3.98 &   3.98 & 6.43  &   6.41 & 10.90 &
  9.33 & 2.93 &   1.38 & 2.94 &   2.91 &TO\\
    & 4909 &  4940 & 3.29 &   3.37 & 9.86 &   8.35 & 1.88 &
  2.73 & 1.30 &   -0.23 & 3.7 &   3.7 & M67\\
\hline
1.36 & 6875 &   6877 & 4.34 &   4.34 & 3.34 &   3.34 & 12.23 &
  19.33 & 3.17 &   3.17 & 2.95  10$^{-2}$ &   3.02 10$^{-2}$ &ZAMS\\
& 6559 &   6521 & 3.97 &   3.95 & 6.63 &   6.74 & 10.81 &
  9.33 & 2.94 &   1.37 & 2.84 &   2.91 &TO\\
& 4709 &   4752 & 2.78 &   3.02 & 26.90 &   22.87 & 1.11 &
  1.60 & 1.26 &   -0.37 & 3.7 &   3.7 & M67\\
\hline
1.37 & 6898 &   6902 & 4.34 &    4.34 & 3.51 &   3.51 & 12.23 &   20.45 & 3.17 &
  3.17 &  3.03 10$^{-2}$ &   2.98 10$^{-2}$ & ZAMS\\
& 6515 &   6555 & 3.94 &   3.95 & 6.95 &   6.88 & 10.40 &   9.87 & 2.93 &   1.15 & 2.88 &   2.82 & TO\\
&  4637 &   4453 & 2.61 &   2.23 & 37.87 &   76.04 & 0.94 &   0.91 & 1.03 &
  -0.61 & 3.7 & 3.7 & M67\\\hline\hline
\end{tabular}
\end{center}
\end{table*} 

\subsection{Evolutionary status and rotation}

Our choice of masses for the models computed are based on the 1.3
M$_\odot$
turn-off mass suggested by Michaud et al. (\cite{Michaud04}) for their 3.7
Gyr best-fitting isochrone to the M67 colour-magnitude diagram, and taking
into account that our sample includes subgiant and giant
stars that should be more massive. Figure~\ref{teff_logg} shows the 
surface gravity as a function of
temperature for our sample stars together with
our models at the age of 3.7 Gyr. Both
types of models, whether they are slow or moderate rotators on the ZAMS, fall
among the data points. The data are well reproduced by rotating models that have passed
the turn-off (see Table~\ref{tab:models}), and that have initial masses
between 1.30 M$_\odot$ and 1.37 M$_\odot$.\\

\begin{figure}[!ht]
   \centering
   \includegraphics[width=9cm]{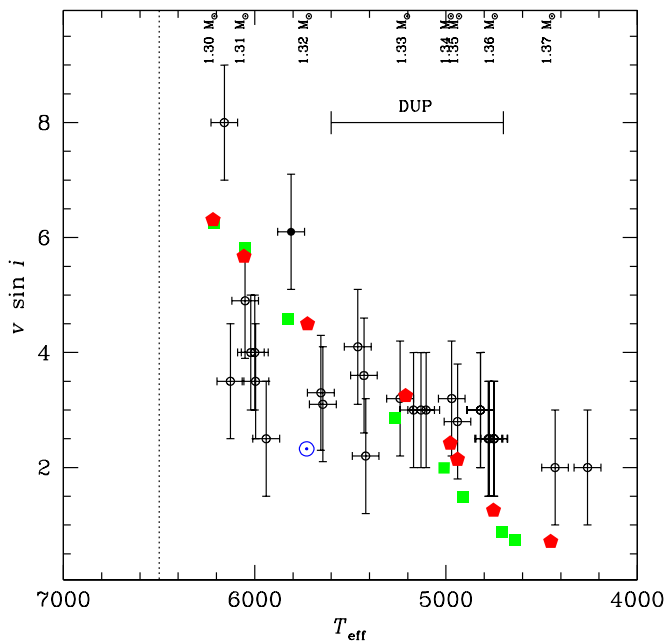}
      \caption{$\upsilon \sin$i versus \Teff~for our sample of stars. 
Symbols are the
      same as in Fig.~\ref{teff_logg}. The theoretical predictions 
here are
      mean \vsini~ and
      correspond to the $\upsilon_{eq}$ values at 3.7 Gyr given in
      Table~\ref{tab:models} multiplied by $\frac{\pi}{4}$. The  filled 
dot is for S1242, which has
      been identified as a peculiar Li-rich subgiant in a previous work of
      ours (Canto Martins et al. \cite{Bruno06}).}
       \label{teff_vsini}
   \end{figure}

Figure~\ref{teff_vsini} presents the rotation velocity
\vsini~ of our data versus temperature and overplotted are the
corresponding  \vsini~ from our models at 3.7 Gyr. The models have an
actual equatorial rotation velocity higher than the values plotted here that were obtained by multiplying the values of Table~\ref{tab:models} by
$\frac{\pi}{4}$ following Chandrasekhar (\cite{Chandrasekhar50}). They 
mimic a mean \vsini~associated to what can be considered a mean equatorial velocity given by
stellar evolution models. The initial velocity and braking parameters 
(when relevant) were chosen to obtain surface velocities between 8 and 1 
\kms~at 3.7
Gyr. For models with $\upsilon_{ZAMS} \simeq 19$ \kms, the 
evolution of the
surface rotation is driven by the radius changes as the star evolves, so
that it essentially drops from the ZAMS to the upper RGB. The more 
massive of our models, which are
also the more evolved, present the lowest surface rotation velocities, in
agreement with the observations. The angular
momentum lost by the wind at the evolutionary phases considered is
accounted for, but is found to be negligible.\\

Concerning the models that undergo
magnetic braking, the figure shows the result of our calibration of the
braking parameter $K$. The extraction of angular momentum when using the 
Kawaler law
is maximum when the model arrives on the ZAMS, because it is proportional
to the surface angular velocity. By the time the model reaches the
turn-off, it is efficiently slowed down and the surface rotation velocity
has dropped by more than a factor of 2 as can be seen in 
Table~\ref{tab:models}. The
evolution of $\upsilon_{\rm surf}$  on the subgiant and red giant 
branches is then dominated by the increase of the radius as for the other family of models.\\

Referring back to Talon \& Charbonnel (\cite{TC03}), we
note that our models at the age of the Hyades have effective
temperatures between 6700 K and 6950 K, and would thus lie on the hot side
of the Li dip. The convective envelope on the main sequence is thus
shallow, and internal gravity waves are not expected to efficiently transport
angular momentum or to influence the transport of chemicals. On the
contrary, based on Palacios et al. (\cite{Palacios03}), we expect
rotational mixing (e.g. meridional circulation and shear induced mixing)
to
efficiently transport Li so that the abundances derived from observations
are reproduced. Let us note that the stars of M67 that are presently
  on the main sequence are less massive than 1.30 M$_\odot$, and probably
  lie on the cool side of the Li dip. Thus according to Talon \&
Charbonnel
  (\cite{TC03}), we should expect an efficient transport of both angular
  momentum and chemicals by internal gravity waves, which should therefore be
  taken into account if these stars were to be modelled.

\subsection{Lithium as tracer of rotational mixing on the main sequence}

\begin{figure}[!t]
   \centering
   \includegraphics[width=9cm]{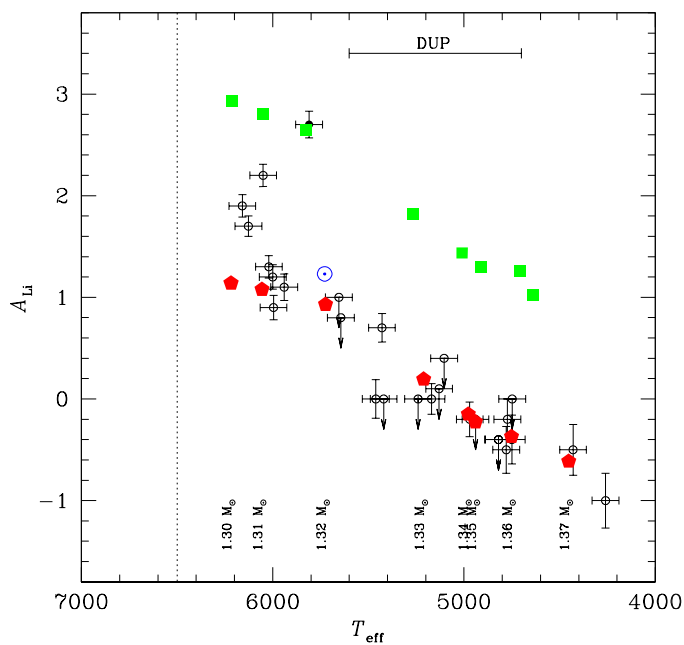}
      \caption{\Ali versus \Teff~for our sample of stars. Symbols are the
      same as in Figs.~\ref{teff_logg} and~\ref{teff_vsini}.}
       \label{li_teff}
   \end{figure}

After chosing the masses and calibrating the initial rotation parameters so
that the gravity and the \vsini~ patterns as a function of effective
temperature are well reproduced for our sample stars, we look at what is
obtained for the lithium abundance. The results are presented in
Fig.~\ref{li_teff}, where we plot the derived lithium abundances
for our M67 sample and the abundances obtained for each of our models at
3.7 Gyr. Li abundances appear here as a clear separator between the
two families of models we considered. Models that are slow rotators and do
not undergo magnetic braking on the ZAMS are ruled out. Actually the Li abundance 
evolves similarly to standard models in these
models and lithium
barely decreases at the surface during the main-sequence evolution, as can
be seen from Table~\ref{tab:models}. On the contrary, models arriving on
the ZAMS with mild rotation (about 19 \kms) and undergoing magnetic
braking experience a reduction of the Li abundance by more than $\approx 1.6$
dex during the main-sequence evolution. Afterwards, when the models 
evolve to the RGB, the 1$^{st}$
DUP further dilutes the Li of the envelope, and it almost disappears at
the
surface as can be seen from the 1.35, 1.36 and 1.37 M$_\odot$ models. 
These models,
presented as filled pentagons on Fig.~\ref{li_teff}, follow the trend
obtained for our M67 sample and nicely reproduce the \Ali~values derived
from observations. \\ 

\begin{figure}[h]
   \centering
   \includegraphics[width=9cm]{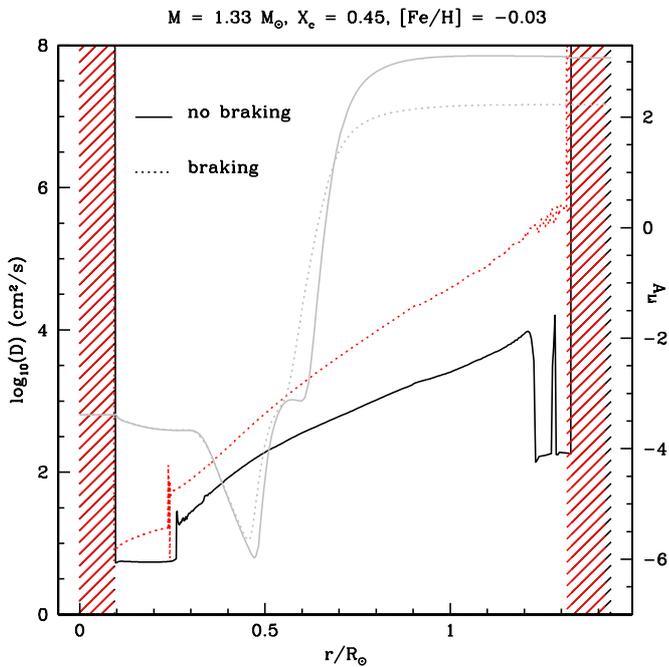}
      \caption{Internal profiles of \Ali (grey) and  of the logarithm of 
the total
      diffusion coefficient (black) as a function of the radial 
coordinate in
      rotating models with initial mass of 1.33 M$_\odot$ with (dotted 
lines)
      and without (solid lines) braking. The profiles are shown for t 
$\simeq$ 1.4 Gyr, when the models are
      on the main sequence (X$_c \simeq 0.46$). Left ordinates refer to 
the scale for the diffusion
      coefficient, while right ordinates represent the scale for the
lithium
      abundance A$_{\rm Li}$. Hatched regions represent convective zones.}
       \label{Dtot_Li}
   \end{figure}

Without braking the angular velocity gradient below the convective envelope is
shallow when the star is on the main sequence. As a consequence, the
shear instability does not develop. Figure~\ref{Dtot_Li} presents the
inner profile of \Ali~ and of the diffusion coefficient for chemicals ($D
=
D_{mer. circ.} + D_{shear} + D_{micro}$)~as a function
of the radial coordinate in the 1.33 M$_\odot$ model while on main 
sequence (t
$\approx 1.4$ Gyr). The solid lines refer to the model without braking,
and we see that the diffusion coefficient is small in the region where Li is
burned and the transport is strongly inefficient. Because the meridional 
circulation velocity
also remains very slow (a few 10$^{-8}$ cm.s$^{-1}$ at most), the 
transport of chemicals is
dominated by the microscopic diffusion, which is a slow process ($D_{\rm  mic} \approx 100~{\rm cm}^2.{\rm s}^{-1}$) 
leading to a small decrease of the surface
lithium abundance, which would also be obtained in a non-rotating model with
microscopic diffusion. On the other hand, the extraction of angular
momentum associated with the magnetic braking leads to an increase of the
angular velocity gradient in the stellar interior in a way that the
shear instability sets in and the meridional circulation becomes stronger
as the star is slowed down on the ZAMS. For the 1.33 M$_\odot$ model, as
can be seen in Fig.~\ref{Dtot_Li}, the diffusion coefficient is two orders
of
magnitude higher than in the previous case just below the convective
envelope, where the meridional circulation velocity is the highest
($\approx 10^{-5}$ cm.s$^{-1}$). It is also more than one order of 
magnitude larger in the region where Li
is destroyed. As a result, as can be seen from the \Ali~profile, lithium
(and hydrogen) diffuses inwards, reducing the gradient around the radial
coordinate 0.65 R$_\odot$, and its surface abundance decreases. The depth
of the DUP is not affected by rotation, and the subsequent decrease of 
\Ali~at
this stage is similar, whether angular momentum was extracted or not during
the early evolution on the main sequence.\\

Let us emphasize that the lithium abundances obtained in our models with 
the adopted
initial velocities and braking parameters fit the upper envelope of the
values derived for the stars in our sample. The abundance determination
lead only to upper limits for stars undergoing the DUP (in the range
T$_{\rm eff}~\in$ [4700 K, 5600 K]). However, we do not expect the 
actual values
to be much lower than these limits considering that for the stars with
T$_{\rm eff}~\approx~ 4500$K in our sample, \Ali~has been measured and is
of about -0.5 dex. This value gives the lower limit for the post-DUP
\Ali~value. Because the shape of the \Ali~vs \Teff~pattern is well reproduced
by
our models, a slight adjustment of the rotation parameters so as to
decrease more the Li on the main sequence would concomitantly bring down
the values for more evolved models.\\

As explained above, we calibrated the braking parameters in our models  
to reproduce the evolution
  of the surface rotation velocity of stars beyond the turn-off in
  M67. This calibration depends on the equatorial rotation velocity
adopted
  on the ZAMS. The models presented in Table~\ref{tab:models} show that
  for $\upsilon_{eq,ZAMS} \simeq 19$ \kms, the amount of angular momentum
  extraction needed to reach \vsini~$\simeq 8.5$ \kms at the turn-off
  following Melo et al. (\cite{Melo01}) induces an internal transport of
  chemicals that results in a decrease of the surface lithium abundance that agrees very well 
  with our observational data. Considering the
  uncertainties on the equatorial rotation velocities of the now giant
  stars of M67 when they arrived on the ZAMS, one may now wonder what
  would happen if we had chosen a faster rotation on the ZAMS. We thus
  computed models of 1.30, 1.33 and 1.36 M$_\odot$ with a larger initial
  content of angular momentum leading to $\upsilon_{eq,ZAMS} \simeq 27.5$
  \kms, and calibrated the magnetic braking parameter of Kawaler's law
  in order to reach \vsini~ values in the observed range when the models
  reach 3.7 Gyr. While the basic stellar parameters and the equatorial
  surface rotation velocities of these faster models resemble those
  presented in Table~\ref{tab:models}, the surface lithium abundance
behaves
very differently and is much lower at 3.7 Gyr. This is the direct
consequence of a stronger shear mixing that develops in the stellar interior
because of the stronger magnetic torque applied at the surface of the star
during its early evolution on the main sequence. This test allows us to
constrain the ZAMS equatorial velocity of stars in the mass range 1.30
M$_\odot$ to 1.37 M$_\odot$ of M67 considering that meridional circulation
and shear turbulence are the main processes transporting lithium in this
stars. We therefore exclude models with ZAMS equatorial velocities higher than
20 \kms.

\section{Conclusions}

We performed a spectroscopic analysis for a unique sample of 27 evolved
stars of the old open cluster M67 following an evolutionary
sequence from the turn-off to the Red Giant Branch.  In this analysis we 
used a spectral
synthesis procedure based on MARCS model atmospheres to derive the stellar
parameters (T$_{eff}$, $\log g$, $[Fe/H]$), microturbulent and rotational
velocities, and the lithium abundance.  The surface lithium content of the analysed stars 
follows a clear evolutionary pattern ranging from the turn-off to the Red 
Giant Branch.
Our results reproduce the lithium abundance dispersion previously 
established for the M67 main-sequence stars and
also confirm the well known gradual decrease of A$_{Li}$ for T$_{eff}$ 
below 5500 K.
The lithium abundances derived for our sample stars clearly indicate 
that this element has been already depleted at the stellar surface during
 the main-sequence stage, which also points to a depletion in excess of 
standard predictions.
The present study  largely extends  the number of evolved stars studied 
in M67. For the first time, it offers solid possibilities for an 
evolutionary study of
different stellar parameters (A$_{Li}$  and \vsini) and also to test non-standard 
transport processes on homogeneous sample of turn-off  and 
evolved stars of M67.\\

 We tested the hypothesis that lithium and surface rotation
  evolution might be connected by comparing our data to the predictions 
of a set of
  stellar evolution models including the transport of matter and angular
  momentum by meridional circulation and shear-induced turbulence. It
  appears that these models (following Maeder \& Zahn \cite{MZ98}
formalism
  for the transport of angular momentum) consistently reproduce
  the entire set of observational data when the evolution of angular
  momentum in the models includes an early phase of magnetic braking
``{\em
  \`a la Kawaler}'', the efficiency of which is calibrated on the \vsini~
  data available for main-sequence and turn-off stars of M67. This result
  is similar to what was also obtained by Smiljanic et al. 
(\cite{Smiljanic10}) for
  the open cluster IC 4651, and confirms the importance of
rotation-induced
  mixing in determining light elements abundance patterns in low-mass
stars
  in particular for those with initial masses larger than 1.30
  M$_\odot$ when at solar metallicity.

\acknowledgements

{This work has been supported by continuous grants from
CNPq Brazilian Agency (J.R. De Medeiros and J.D. do Nascimento Jr.) and by 
financial support from the french CNRS/INSU Programme National de Physique 
Stellaire (A. L\`ebre, A. Palacios, O. Richard, P. De Laverny). B. L. Canto Martins 
acknowledges CAPES Brazilian Agency for a Ph.D. fellowship. The authors warmly 
thank Bertrand Plez and Thomas Masseron for their help on the use of MARCS 
models and {\it TurboSpectrum} routines, as well as for the molecular lists 
they provided.
 }

\Online 

\begin{table}[!htp]
\centering
\caption{Iron line list used for curve of growth and spectral synthesis, with atomic parameters, corrected oscillator strengths and solar and Arcturus equivalent width measurements.}
\label{Felines}
\begin{tabular}{lccccc}
\hline \hline
Element &	$\lambda$	&	$\chi_{exc}$	&	$\log gf$	& $EW_\odot $ & $EW_{Arc} $	\\
 &	(\AA)	&	(eV)	&		& (m\AA) & (m\AA)	\\
\hline \hline
Fe I	&	5036.919	&	3.017	&	-2.938	&	24.3	&	64.8	\\
Fe I	&	5044.211	&	2.851	&	-2.128	&	70.3	&	117.8	\\
Fe I	&	5054.643	&	3.640	&	-2.051	&	38.7	&	65.0	\\
Fe I	&	5228.377	&	4.220	&	-1.115	&	57.5	&	83.8	\\
Fe I	&	5242.493	&	3.634	&	-1.007	&	81.2	&	112.8	\\
Fe I	&	5247.051	&	0.087	&	-4.936	&	66.0	&	175.6	\\
Fe I	&	5321.108	&	4.434	&	-1.301	&	40.0	&	59.2	\\
Fe I	&	5322.041	&	2.279	&	-2.953	&	60.7	&	115.1	\\
Fe I	&	5326.143	&	3.573	&	-2.211	&	35.8	&	69.9	\\
Fe I	&	5373.709	&	4.473	&	-0.830	&	60.7	&	76.9	\\
Fe I	&	5386.330	&	4.154	&	-1.740	&	31.1	&	51.3	\\
Fe I	&	5522.444	&	4.209	&	-1.450	&	43.4	&	62.0	\\
Fe I	&	5543.936	&	4.217	&	-1.080	&	62.0	&	81.8	\\
Fe I	&	5618.633	&	4.209	&	-1.316	&	50.5	&	69.5	\\
Fe I	&	5638.262	&	4.220	&	-0.810	&	75.1	&	97.3	\\
Fe I	&	5701.547	&	2.559	&	-2.276	&	82.3	&	142.4	\\
Fe I	&	5705.465	&	4.301	&	-1.492	&	38.4	&	57.8	\\
Fe I	&	5741.848	&	4.256	&	-1.674	&	32.0	&	51.5	\\
Fe I	&	5775.081	&	4.220	&	-1.188	&	59.2	&	81.1	\\
Fe I	&	5778.453	&	2.588	&	-3.510	&	23.0	&	73.0	\\
Fe I	&	5806.725	&	4.607	&	-0.975	&	52.0	&	68.1	\\
Fe I	&	5811.914	&	4.143	&	-2.390	&	10.7	&	23.5	\\
Fe I	&	5852.219	&	4.548	&	-1.260	&	39.4	&	---	\\
Fe I	&	5853.148	&	1.485	&	-5.200	&	7.3	&	62.8	\\
Fe I	&	5855.077	&	4.608	&	-1.583	&	21.5	&	34.1	\\
Fe I	&	5856.088	&	4.294	&	-1.615	&	33.5	&	54.5	\\
Fe I	&	5858.778	&	4.220	&	-2.245	&	12.6	&	27.2	\\
Fe I	&	5916.247	&	2.453	&	-2.994	&	55.2	&	112.9	\\
Fe I	&	5927.789	&	4.652	&	-1.105	&	41.7	&	54.1	\\
Fe I	&	5934.657	&	3.928	&	-1.225	&	71.9	&	101.8	\\
Fe I	&	5956.694	&	0.859	&	-4.630	&	52.5	&	148.9	\\
Fe I	&	5976.777	&	3.943	&	-1.365	&	65.0	&	101.8	\\
Fe I	&	5987.066	&	4.795	&	-0.556	&	64.1	&	79.2	\\
Fe I	&	6003.014	&	3.881	&	-1.140	&	79.5	&	107.8	\\
Fe I	&	6019.362	&	3.573	&	-3.280	&	5.2	&	17.8	\\
Fe I	&	6027.051	&	4.076	&	-1.190	&	9.4	&	89.2	\\
Fe I	&	6054.070	&	4.371	&	-2.245	&	9.4	&	17.9	\\
Fe I	&	6056.005	&	4.733	&	-0.490	&	69.7	&	81.7	\\
Fe I	&	6079.009	&	4.652	&	-1.050	&	45.7	&	---	\\
Fe I	&	6105.128	&	4.548	&	-2.000	&	11.1	&	21.2	\\
Fe I	&	6120.249	&	0.915	&	-5.910	&	5.3	&	66.7	\\
Fe I	&	6151.618	&	2.176	&	-3.359	&	50.6	&	113.9	\\
Fe I	&	6157.728	&	4.076	&	-1.270	&	61.7	&	92.7	\\
Fe I	&	6159.373	&	4.607	&	-1.920	&	12.0	&	---	\\
Fe I	&	6165.358	&	4.143	&	-1.535	&	44.9	&	66.9	\\
Fe I	&	6180.204	&	2.727	&	-2.736	&	54.7	&	108.5	\\
Fe I	&	6187.988	&	3.943	&	-1.735	&	46.3	&	74.2	\\
Fe I	&	6226.736	&	3.883	&	-2.145	&	28.1	&	---	\\
Fe I	&	6229.226	&	2.845	&	-2.970	&	37.2	&	86.0	\\
Fe I	&	6240.646	&	2.223	&	-3.353	&	48.9	&	---	\\
Fe I	&	6265.136	&	2.176	&	-2.700	&	83.2	&	160.2	\\
Fe I	&	6270.225	&	2.858	&	-2.670	&	51.9	&	102.2	\\
Fe I	&	6271.277	&	3.332	&	-2.776	&	23.4	&	58.3	\\
Fe I	&	6297.795	&	2.223	&	-2.871	&	73.4	&	143.3	\\
Fe I	&	6315.809	&	4.076	&	-1.683	&	40.9	&	66.9	\\
Fe I	&	6380.743	&	4.186	&	-1.396	&	50.6	&	---	\\
Fe I	&	6392.539	&	2.279	&	-3.990	&	17.5	&	---	\\
Fe I	&	6498.939	&	0.958	&	-4.701	&	46.1	&	144.3	\\
Fe I	&	6574.228	&	0.990	&	-4.940	&	33.1	&	125.7	\\
Fe I	&	6575.018	&	2.588	&	-2.765	&	64.6	&	---	\\
Fe I	&	6581.208	&	1.485	&	-4.730	&	20.2	&	97.5	\\
Fe I	&	6608.026	&	2.279	&	-4.010	&	17.3	&	74.5	\\
Fe I	&	6627.543	&	4.548	&	-1.560	&	27.2	&	42.1	\\
Fe I	&	6646.932	&	2.608	&	-4.015	&	9.3	&	50.0	\\
Fe I	&	6653.851	&	4.154	&	-2.470	&	9.9	&	23.6	\\
Fe I	&	6699.140	&	4.593	&	-2.179	&	8.3	&	15.2	\\
\hline \hline
\end{tabular}
\end{table}

\begin{table}[!htp]
\begin{flushleft}
{\bf Table 6.} Cont.
\end{flushleft}
\centering
\begin{tabular}{lccccc}
\hline \hline
Element &	$\lambda$	&	$\chi_{exc}$	&	$\log gf$	& $EW_\odot $ & $EW_{Arc} $	\\
 &	(\AA)	&	(eV)	&		& (m\AA) & (m\AA)	\\
\hline \hline
Fe I	&	6703.568	&	2.758	&	-3.080	&	38.5	&	89.5	\\
Fe I	&	6704.480	&	4.217	&	-2.650	&	6.4	&	14.0	\\
Fe I	&	6705.102	&	4.607	&	-1.136	&	46.3	&	61.0	\\
Fe I	&	6707.420	&	4.610	&	-2.245	&	6.3	&	---	\\
Fe I	&	6710.318	&	1.485	&	-4.865	&	16.3	&	89.2	\\
Fe I	&	6713.045	&	4.607	&	-1.553	&	24.2	&	---	\\
Fe I	&	6713.190	&	4.143	&	-2.475	&	10.2	&	21.3	\\
Fe I	&	6713.740	&	4.795	&	-1.465	&	21.4	&	28.4	\\
Fe I	&	6715.382	&	4.607	&	-1.535	&	26.5	&	---	\\
Fe I	&	6716.233	&	4.580	&	-1.875	&	15.3	&	---	\\
Fe I	&	6725.357	&	4.103	&	-2.250	&	17.7	&	35.4	\\
Fe I	&	6726.665	&	4.607	&	-1.129	&	47.1	&	59.9	\\
Fe I	&	6733.149	&	4.638	&	-1.485	&	26.9	&	38.2	\\
Fe I	&	6739.520	&	1.557	&	-4.929	&	12.4	&	76.9	\\
Fe I	&	6752.705	&	4.638	&	-1.314	&	35.7	&	54.7	\\
Fe I	&	6786.858	&	4.191	&	-1.950	&	25.3	&	44.7	\\
Fe I	&	6806.845	&	2.727	&	-3.167	&	35.2	&	89.7	\\
Fe I	&	6810.261	&	4.607	&	-1.068	&	49.0	&	65.1	\\
Fe I	&	6971.932	&	3.018	&	-3.475	&	12.8	&	47.4	\\
Fe I	&	7112.167	&	2.990	&	-3.035	&	31.6	&	31.6	\\
Fe I	&	7189.150	&	3.071	&	-2.761	&	36.5	&	36.5	\\
Fe I	&	7401.680	&	4.186	&	-1.625	&	39.9	&	39.9	\\
Fe I	&	7710.362	&	4.220	&	-1.231	&	62.2	&	62.2	\\
Fe I	&	7723.204	&	2.279	&	-3.542	&	37.8	&	37.8	\\
Fe I	&	7941.087	&	3.274	&	-2.550	&	41.9	&	41.9	\\
Fe II	&	4993.354	&	2.807	&	-3.772	&	38.0	&	---	\\
Fe II	&	5100.627	&	2.807	&	-4.260	&	17.3	&	37.4	\\
Fe II	&	5132.664	&	2.807	&	-4.130	&	21.5	&	16.8	\\
Fe II	&	5136.794	&	2.844	&	-4.440	&	10.0	&	---	\\
Fe II	&	5197.570	&	3.230	&	-2.450	&	74.5	&	71.1	\\
Fe II	&	5234.625	&	3.221	&	-2.390	&	78.3	&	78.8	\\
Fe II	&	5264.805	&	3.230	&	-3.205	&	44.9	&	38.8	\\
Fe II	&	5325.553	&	3.221	&	-3.310	&	42.0	&	38.6	\\
Fe II	&	5414.070	&	3.221	&	-3.677	&	25.5	&	22.5	\\
Fe II	&	6084.103	&	3.199	&	-3.808	&	20.3	&	20.9	\\
Fe II	&	6369.458	&	2.891	&	-4.253	&	18.1	&	19.2	\\
Fe II	&	6416.923	&	3.892	&	-2.790	&	38.1	&	35.4	\\
Fe II	&	6456.383	&	3.903	&	-2.075	&	60.6	&	46.7	\\
Fe II	&	7224.475	&	3.889	&	-3.317	&	19.1	&	19.1	\\
\hline \hline
\end{tabular}
\end{table}

\end{document}